\documentclass[usenatbib]{mn2e}
\usepackage[T1]{fontenc}
\usepackage{ae,aecompl}

\usepackage{graphicx}   

\newcommand{\JHK}{JHK_{\rm s}}
\newcommand{\Ks}{K_{\mathrm{s}}}

\newcommand{\AK}{A_{\Ks}}

\newcommand{\EHK}{E_{H-\Ks}}
\newcommand{\AKEHK}{A_{\Ks}/E_{H-\Ks}}

\title[A lack of Cepheids in the inner Galaxy]
{A lack of classical Cepheids in the inner part of\\the Galactic disk}
\author[N. Matsunaga~{et~al.}]
{Noriyuki Matsunaga$^{1}$\thanks{E-mail:matsunaga@astron.s.u-tokyo.ac.jp},
Michael W. Feast$^{2,3}$,
Giuseppe Bono$^{4,5}$,
Naoto Kobayashi$^{6,7,8}$, 
\and
Laura Inno$^{9}$,
Takahiro Nagayama$^{10}$,
Shogo Nishiyama$^{11}$,
Yoshiki Matsuoka$^{12}$,
\and
Tetsuya Nagata$^{13}$
\\
$^{1}$ Department of Astronomy, The University of Tokyo, 7-3-1 Hongo, Bunkyo-ku, Tokyo 113-0033, Japan \\
$^{2}$ Astrophysics, Cosmology and Gravity Centre, Astronomy Department, University of Cape Town, Rondebosch, 7701, South Africa\\
$^{3}$ South African Astronomical Observatory, PO Box 9, Observatory 7935, South Africa \\
$^{4}$ Dipartimento di Fisica, Universit\'{a} di Roma Tor Vergata, Via della Ricerca Scientifica 1, 00133 Rome, Italy \\
$^{5}$ Instituto Nazionale de Astrofisica, Osservatorio Astronomico di Roma, Via Frascati 33, 00040 Monte Porzio Catone, Italy \\
$^{6}$ Institute of Astronomy, School of Science, The University of Tokyo, 2-21-1 Osawa, Mitaka, Tokyo 181-0015, Japan \\
$^{7}$ Kiso Observatory, Institute of Astronomy, School of Science, The University of Tokyo, 10762-30 Mitake, \\
~Kiso-machi, Kiso-gun, Nagano 397-0101, Japan \\
$^{8}$ Laboratory of Infrared High-resolution spectroscopy (LiH), Koyama Astronomical Observatory, Kyoto Sangyo University, \\
~Motoyama, Kamigamo, Kita-ku, Kyoto-603-8555, Japan \\
$^{9}$ Max Planck Institute for Astronomy, K\"{o}nigstuhl 17, D-69117 Heidelberg, Germany \\
$^{10}$ Department of Physics and Astronomy, Kagoshima University, 1-21-35 Korimoto, Kagoshima, 890-0065, Japan \\
$^{11}$ Miyagi University of Education, Aoba-ku, Sendai, Miyagi 980-0845, Japan \\
$^{12}$ National Astronomical Observatory of Japan, 2-21-1 Osawa, Mitaka, Tokyo 181-8588, Japan \\
$^{13}$ Department of Astronomy, Kyoto University, Kitashirakawa-Oiwake-cho, Sakyo-ku, Kyoto 606-8502, Japan 
}

\date{Accepted 2016 June 22. Received 2016 June 22; in original form 2016 May 21}

\pubyear{2016}

\begin{document}
\label{firstpage}
\pagerange{\pageref{firstpage}--\pageref{lastpage}} \pubyear{2016}
\maketitle

\begin{abstract}
Recent large-scale infrared surveys have been revealing
stellar populations in the inner Galaxy seen through
strong interstellar extinction in the disk.
In particular, classical Cepheids with their period--luminosity
and period--age relations are useful tracers of Galactic structure
and evolution.
Interesting groups of Cepheids reported recently include
four Cepheids in the Nuclear Stellar Disk (NSD), about 200~pc around
the Galactic Centre,
found by Matsunaga {et~al.} and those spread across the inner part of the disk
reported by D\'{e}k\'{a}ny and collaborators.
We here report our discovery of nearly thirty classical Cepheids towards
the bulge region, some of which are common with D\'{e}k\'{a}ny {et~al.},
and discuss the large impact of the reddening correction
on distance estimates for these objects.
Assuming that the four Cepheids in the NSD
are located at the distance of the Galactic Centre
and that the near-infrared extinction law, 
i.e.\  wavelength dependency of the interstellar extinction, is 
not systematically different between the NSD and other bulge lines-of-sight,
most of the other Cepheids presented here are located significantly
further than the Galactic Centre. This suggests a lack of Cepheids
in the inner 2.5~kpc region of the Galactic disk except the NSD.
Recent radio observations show
a similar distribution of star-forming regions.
\end{abstract}


\section{Introduction}
\label{sec:Intro}

The structure of the inner Galaxy still remains to be revealed.
While recent infrared surveys have led to 
discoveries of new features in the bulge
\citep[e.g.][]{Nishiyama-2005,Saito-2011},
the distribution of young stars, say younger than 1~Gyr, restricted to
within a degree of the Galactic plane is still elusive.
In addition to the strong foreground extinction towards the inner Galaxy,
the high density of older stars in the bulge makes it difficult to
trace the younger component(s).

It is therefore required to focus on some prominent tracers
whose ages are easily deduced.
For example, the Nuclear Stellar Disk (hereinafter NSD)
has been long known to host
massive stellar clusters whose ages are several Myr
(Figer, McLean \& Morris \citeyear{Figer-1999}; 
\citealt{Figer-2002}). This stellar disk extends
$\sim$400~pc around Sgr~A$^*$ and co-exists with
the massive reservoir of interstellar matter known as
the Central Molecular Zone
(Launhardt, Zylka \& Mezger \citeyear{Launhardt-2002}).
This region also hosts current star formation and young stellar objects
spread across the NSD \citep{YusefZadeh-2009,Mauerhan-2010}.

Classical Cepheids are also useful tracers of young stellar populations.
They are pulsating supergiants with the famous
period--luminosity relation (PLR) which enables us to estimate
their distances reasonably well. They are evolved from
intermediate-mass stars, 3.5--11~M$_\odot$, and their period--age relation
allows us to trace stellar populations aged 10--300~Myr \citep{Bono-2005}.
\citet{Matsunaga-2011} discovered three classical Cepheids in the NSD.
The three Cepheids have similar periods,
$\sim$20~days, which indicated that they are around 25~Myr old.
\cite{Matsunaga-2015}, in addition to reporting another
Cepheid with very similar characteristics, provided further support for
the membership of the four Cepheids to the NSD
based on radial velocities from near-IR high-resolution spectra.

\citet{Dekany-2015a,Dekany-2015b} recently reported dozens of
new Cepheids found in the VVV survey
(VISTA Variables in the V\'{i}a L\'{a}ctea Survey, \citealt{Minniti-2010}).
Their new Cepheids are located towards the bulge region
and more spread than the NSD, and \citet{Dekany-2015b} 
concluded that they are in the inner part of 
the disk, within a few kpc around Sgr~A$^{*}$.
Radio observations, however, do not support
such ubiquitous distribution of young objects in the inner part of the disk.
Recent deep radio observations have discovered two kinds of objects
that trace star formation activities in the inner Galaxy:
H$_{\rm II}$ regions detected in recombination line and
radio continuum emission \citep[e.g.][]{Anderson-2011,Anderson-2012},
and maser spots in high-mass star-forming regions \citep[e.g.][]{Sanna-2014}.
H$_{\rm I}$ absorption seen in the radio continuum of the former
makes it possible to estimate locations of the H$_{\rm II}$ regions,
while Very Long Baseline Interferometer observations of
the latter, if conditions allow, enable one to obtain trigonometric distances. 
As discussed by \citet{Jones-2013}, no H$_{\rm II}$ regions have
been identified within 4~kpc around Sgr~A$^*$ except in the NSD,
while at the edge of this void exist
the expanding 3~kpc arms \citep{Dame-2008}
with which H$_{\rm II}$ regions are associated.

In this paper, we report classical Cepheids newly discovered
in our near-IR survey and discuss their distribution.
In particular, we discuss the large effect of the extinction law.
The ratio of total-to-selective extinction $\AKEHK$ is important,
in our discussions, for estimating the extinction in $\Ks$ from
the reddening in $H-\Ks$.
\citet{Nishiyama-2006} obtained $\AKEHK = 1.44 \pm 0.01$
based on their photometric data for red clump towards the bulge,
while \citet{Nishiyama-2009} combined the 2MASS $\Ks$ magnitudes
of red clump stars with the 2MASS colours of red giant
stars in the same areas to obtain $\AKEHK = 1.61 \pm 0.04$.
This difference leads to estimates
of extinction differing by $\sim$10~\%; a smaller $\AKEHK$ value
would result in a larger estimate of distance.
Objects towards the Galactic Centre have large reddenings,
1.5--2.0~mag in $\EHK$, thus leading to a large difference
in distance moduli, 0.3--0.4~mag,
which corresponds to 1.0--1.5~kpc at the distance of $\sim$8.0~kpc.
\citet{Dekany-2015a,Dekany-2015b} adopted the $\AKEHK$ value
from \citet{Nishiyama-2009} after a transformation into
their VVV photometric system, while adopting the smaller value from
\citet{Nishiyama-2006} would put their Cepheids further than
they estimated as we discuss below in more detail.

\section{Observation and data analysis}
\label{sec:Observation}

Our observations were conducted using the IRSF 1.4~m telescope and
the SIRIUS camera (\citealt{Nagashima-1999}; \citealt{Nagayama-2003})
which collects images in the $J$, $H$ and $\Ks$ bands, simultaneously.
These observations are thus on the same colour system as those of
\citet{Nishiyama-2006}.
The observed field is composed of 142 fields-of-view of IRSF/SIRIUS,
covering $\sim$2.3~deg$^2$,
between $-10^\circ$ and $+10^\circ$ in Galactic longitude
along the Galactic plane, i.e.~$0^\circ$ in Galactic latitude.
Observations at about 30 epochs were made between 2007 and 2012.

The basic data analysis was done in the same manner as in
\citet{Matsunaga-2009,Matsunaga-2013}.
In short, point-spread-function (PSF) fitting photometry was performed
on every image using the {\small IRAF/DAOPHOT} package,
and variable stars were searched for by considering
the standard deviations of time-series photometric data.
The variability search was done using the three-band datasets independently,
so that we could find variables even if they are visible only in
one of the $\JHK$ bands.
In addition, we included the variable stars in \citet{Matsunaga-2013}
which are within the fields of this survey even if their variations
are slightly smaller than the detection limits of variability.
This survey is slightly shallower than that of \citet{Matsunaga-2013},
and as a result many objects are too faint to be detected in $J$ band.

The saturation limits are 9.0, 9.0 and 8.5~mag in $J$, $H$ and $\Ks$,
respectively.
On the other hand, the detection limits vary across the survey region depending
on the crowding. For example,
the detection limits are 15.2, 14.9, and 14.1~mag
for relatively sparse regions, $|l|>8^\circ$,
while for crowded regions, $|l|<1^\circ$,
they get shallower, 15.0, 14.1, 13.0~mag in $\JHK$.
The definition of these limits is described in \citet{Matsunaga-2009}.

In this paper we use only $H$- and $\Ks$-band
magnitudes for estimating distances to Cepheids as done by \citet{Dekany-2015a,Dekany-2015b}. We adopt the PLR of 
classical Cepheids from \citet{Matsunaga-2013},
\begin{eqnarray}
M(H)=-3.256 (\log P -1.3) -6.562 \label{eq:PHR} \\
M(\Ks)=-3.295 (\log P -1.3) -6.685 \label{eq:PKR} 
\end{eqnarray}
These relations were calibrated based on
{\it Hubble Space Telescope} parallaxes
of Cepheids in the solar neighbourhood \citep{Benedict-2007,vanLeeuwen-2007}.
These PLRs and the extinction coefficient of $\AKEHK = 1.44$
\citep{Nishiyama-2006} are combined with observed magnitudes, $H$ and $\Ks$,
to estimate
the distance modulus $\mu_0$ and the foreground extinction $\AK$.

\section{Results}
\label{sec:results}

We detected approximately 100 variable stars with period between 0.1 and 60~d,
among which we identified 29 classical Cepheids
(Table~\ref{tab:catalogue}).
We did not always detect the variables in all of the $\JHK$ bands;
Table~\ref{tab:catalogue} includes $M$flag which we also used
in \citet{Matsunaga-2013,Matsunaga-2015}
to show the reasons of non-detection or the qualities
of the listed magnitudes.
All the objects we report here were detected in
both $H$ and $\Ks$ bands, while roughly half of them were
too faint in the $J$ band.
Table~\ref{tab:catalogue} also lists 
$\AK$ and $\mu_0$ derived with $H$- and $\Ks$-band mean magnitudes.

\begin{table*}
\begin{minipage}{170mm}
\caption{
Catalogue of classical Cepheids detected in our survey.
After the numbering ID, the ID combining
RA and Dec.~(J2000.0) follows. Then listed are
$\JHK$ mean magnitudes,
amplitudes, $M$flag, periods,
distance moduli $\mu_0$ and extinctions $\AK$.
The mean magnitudes are intensity-scale means of maximum
and minimum, and the amplitudes refer to peak-to-valley variations.
The definition of the $M$flag is given in
\citet{Matsunaga-2009}.
\label{tab:catalogue}}
\begin{center}
\begin{tabular}{rcrrrrrrccrr}
\hline
\multicolumn{1}{c}{No.} & ID & \multicolumn{1}{c}{$J$} & \multicolumn{1}{c}{$H$} & \multicolumn{1}{c}{$\Ks$} & \multicolumn{1}{c}{$\Delta J$} & \multicolumn{1}{c}{$\Delta H$} & \multicolumn{1}{c}{$\Delta \Ks$} & $M$flag & Period & \multicolumn{1}{c}{$\AK$} & \multicolumn{1}{c}{$\mu_0$} \\
 & & \multicolumn{1}{c}{(mag)} & \multicolumn{1}{c}{(mag)} & \multicolumn{1}{c}{(mag)} & \multicolumn{1}{c}{(mag)} & \multicolumn{1}{c}{(mag)} & \multicolumn{1}{c}{(mag)} &  & (d) & \multicolumn{1}{c}{(mag)} & \multicolumn{1}{c}{(mag)} \\
\hline
 1 & 17201462$-$3711160 & 15.21 & 13.19 & 12.03 & 0.67 & 0.28 & 0.25 & 000 & 5.0999 & 1.53 & 15.24 \\
 2 & 17201826$-$3658528 & 15.22 & 12.83 & 11.54 & 0.38 & 0.27 & 0.18 & 000 & 9.944 & 1.70 & 15.53 \\
 3 & 17241258$-$3601469 & 13.66 & 11.58 & 10.46 & 0.38 & 0.34 & 0.35 & 000 & 17.597 & 1.44 & 15.53 \\
 4 & 17263471$-$3516241 &  \multicolumn{1}{c}{---}  & 13.65 & 11.80 &  \multicolumn{1}{c}{---}  & 0.35 & 0.27 & 300 & 13.405 & 2.50 & 15.42 \\
 5 & 17265423$-$3501081 &  \multicolumn{1}{c}{---}  & 15.98 & 13.99 &  \multicolumn{1}{c}{---}  & 0.67 & 0.37 & 360 & 4.2904 & 2.73 & 15.75 \\
 6 & 17295917$-$3409551 & 12.30 & 10.30 &  9.20 & 0.51 & 0.48 & 0.42 & 000 & 31.52 & 1.40 & 15.14 \\
 7 & 17321407$-$3323595 & 14.05 & 12.09 & 10.98 & 0.33 & 0.21 & 0.21 & 000 & 9.907 & 1.44 & 15.22 \\
 8 & 17384614$-$3126228 &  \multicolumn{1}{c}{---}  & 12.31 & 10.77 &  \multicolumn{1}{c}{---}  & 0.40 & 0.42 & 300 & 25.03 & 2.03 & 15.74 \\
 9 & 17404103$-$3041386 & 15.16 & 12.08 & 10.47 & 0.45 & 0.49 & 0.42 & 600 & 23.88 & 2.14 & 15.28 \\
10 & 17445691$-$2913338 & 14.95 & 12.14 & 10.32 & 1.57 & 0.43 & 0.36 & 600 & 18.87 & 2.45 & 14.48 \\
11 & 17453089$-$2903106 &  \multicolumn{1}{c}{---}  & 12.39 & 10.30 &  \multicolumn{1}{c}{---}  & 0.48 & 0.40 & 300 & 22.75 & 2.83 & 14.34 \\
12 & 17453227$-$2902553 & 15.30 & 11.96 & 10.10 & 0.57 & 0.45 & 0.53 & 000 & 19.96 & 2.50 & 14.28 \\
13 & 17460602$-$2846551 & 15.63 & 12.02 & 10.14 & 0.46 & 0.51 & 0.42 & 300 & 23.52 & 2.53 & 14.53 \\
14 & 17494143$-$2727145 &  \multicolumn{1}{c}{---}  & 12.92 & 11.51 &  \multicolumn{1}{c}{---}  & 0.31 & 0.25 & 300 & 10.48 & 1.87 & 15.40 \\
15 & 17501126$-$2719429 &  \multicolumn{1}{c}{---}  & 12.58 & 10.95 &  \multicolumn{1}{c}{---}  & 0.50 & 0.42 & 300 & 16.52 & 2.17 & 15.19 \\
16 & 17503049$-$2713466 &  \multicolumn{1}{c}{---}  & 14.47 & 12.23 &  \multicolumn{1}{c}{---}  & 0.54 & 0.34 & 300 & 12.655 & 3.06 & 15.20 \\
17 & 17511376$-$2648559 &  \multicolumn{1}{c}{---}  & 13.95 & 12.13 &  \multicolumn{1}{c}{---}  & 0.35 & 0.28 & 300 & 12.9488 & 2.45 & 15.74 \\
18 & 17522166$-$2631194 &  \multicolumn{1}{c}{---}  & 14.25 & 12.27 &  \multicolumn{1}{c}{---}  & 0.35 & 0.27 & 300 & 11.9921 & 2.69 & 15.54 \\
19 & 17522894$-$2623400 & 15.55 & 12.47 & 10.91 & 0.64 & 0.44 & 0.41 & 600 & 22.63 & 2.07 & 15.71 \\
20 & 17523814$-$2619433 & 14.01 & 11.22 &  9.86 & 0.55 & 0.52 & 0.43 & 000 & 38.16 & 1.77 & 15.71 \\
21 & 17544025$-$2534395 &  \multicolumn{1}{c}{---}  & 13.40 & 11.47 &  \multicolumn{1}{c}{---}  & 0.47 & 0.44 & 300 & 17.162 & 2.61 & 15.33 \\
22 & 17573141$-$2430267 &  \multicolumn{1}{c}{---}  & 13.81 & 11.74 &  \multicolumn{1}{c}{---}  & 0.51 & 0.47 & 300 & 24.32 & 2.80 & 15.91 \\
23 & 18012448$-$2254446 &  \multicolumn{1}{c}{---}  & 14.79 & 12.75 &  \multicolumn{1}{c}{---}  & 0.28 & 0.22 & 360 & 11.2397 & 2.77 & 15.84 \\
24 & 18012508$-$2254283 &  \multicolumn{1}{c}{---}  & 14.85 & 12.73 &  \multicolumn{1}{c}{---}  & 0.56 & 0.32 & 360 & 11.2157 & 2.89 & 15.70 \\
25 & 18021484$-$2227107 &  \multicolumn{1}{c}{---}  & 13.90 & 11.99 &  \multicolumn{1}{c}{---}  & 0.44 & 0.41 & 300 & 17.12 & 2.58 & 15.88 \\
26 & 18032993$-$2203225 & 14.55 & 11.68 & 10.35 & 0.70 & 0.48 & 0.46 & 000 & 39.68 & 1.72 & 16.30 \\
27 & 18035395$-$2158117 & 15.42 & 11.91 & 10.23 & 0.76 & 0.41 & 0.42 & 600 & 45.3 & 2.22 & 15.87 \\
28 & 18055284$-$2106419 &  \multicolumn{1}{c}{---}  & 13.10 & 11.44 &  \multicolumn{1}{c}{---}  & 0.49 & 0.38 & 300 & 21.41 & 2.21 & 16.01 \\
29 & 18070782$-$2034501 &  \multicolumn{1}{c}{---}  & 12.74 & 10.94 &  \multicolumn{1}{c}{---}  & 0.37 & 0.38 & 300 & 19 & 2.42 & 15.14 \\
\hline
\end{tabular}
\end{center}
\end{minipage}
\end{table*}

The short-period variables we detected include type II Cepheids
and a few other types of variables like eclipsing binaries.
We will describe details of the classification and
results for other types in a forthcoming paper.
The method of the classification is briefly given
in \citet{Matsunaga-2014}, but an important step is
comparing their estimated distances and extinctions with
a three-dimensional extinction map.
With a given set of magnitude and colour,
two types of Cepheids (classical and type II) would lead to
significantly different distances, $D$, by a factor of $\sim$2, while 
the estimate of $\AK$ is not largely affected thanks to
the similarity of their period--colour relations.
Thus, $(D, \AK)$ in comparison with an extinction map along
the line of sight allows us to determine the Cepheid type.
We used the three-dimensional extinction map by \citet{Schultheis-2014}. 
\citet{Dekany-2015a,Dekany-2015b} used a similar method
to discriminate classical Cepheids from a large number of type II Cepheids
in the bulge.

The extinction map depends on the extinction coefficient.
\citet{Schultheis-2014} used the extinction coefficient from
\citet{Nishiyama-2009} as done by D\'{e}k\'{a}ny {et~al.},
but their $\AKEHK$ is 1.63 considering the transformation from
the 2MASS to the VVV system. The effect of this transformation,
0.02 in $\AKEHK$, is negligible in the following discussions.
Let us assume that an observed feature in the $(H-\Ks)$-$\Ks$ diagram
gives $\AK = 1.61$ at the distance of 8~kpc with $\AKEHK=1.61$ adopted,
i.e.~the observed reddening is 1~mag. If we adopted $\AKEHK=1.44$,
the same feature would have $\AK = 1.44$~mag at the distance of 8.6~kpc.
We transformed the extinction map by \citet{Schultheis-2014} in this way,
which typically reduced the $\AK$ value of the original map
by $\sim$10~\% at each distance, and then compared the $(D, \AK)$ curve with
the parameters of our Cepheids. Such a change in the extinction map,
however, did not affect our conclusion on Cepheid types
because the predicted distances for the two types of Cepheids
are significantly different compared to the effect of
the extinction coefficient.

\section{Discussion}
\label{sec:discussion}

\subsection{Comparison with Cepheids in D\'{e}k\'{a}ny {et~al.}}

D\'{e}k\'{a}ny {et~al.} discovered 
37 classical Cepheids from VVV (two in \citeyear{Dekany-2015a}
and 35 in \citeyear{Dekany-2015b}).
Among these objects, 11 are located within our survey region.
Table~\ref{tab:match} lists our objects which have counterparts
in their catalogue together with the distances and reddenings they derived.
All of them were detected in our survey although some of their light curves
are rather noisy.
Comparing $\AK$ and $\mu_0$ in Tables~\ref{tab:catalogue} and \ref{tab:match},
we notice that significant differences, 0.25--0.7~mag, are present
between the results in \citet{Dekany-2015a,Dekany-2015b} and ours.

\begin{table}
\begin{minipage}{80mm}
\caption{
Matches in previous catalogs (a---\citealt{Dekany-2015a}, b---\citealt{Dekany-2015b}).
The reference values of periods and $\JHK$ magnitudes are listed when available.
\label{tab:match}}
\begin{center}
\begin{tabular}{rccrrrrrrrr}
\hline
\multicolumn{1}{c}{No.} & Ref & ID & \multicolumn{1}{c}{$J$} & \multicolumn{1}{c}{$H$} & \multicolumn{1}{c}{$\Ks$} & \multicolumn{1}{c}{$\AK$} & \multicolumn{1}{c}{$\mu _0$} \\ 
 & & & \multicolumn{1}{c}{(mag)} & \multicolumn{1}{c}{(mag)} & \multicolumn{1}{c}{(mag)} & \multicolumn{1}{c}{(mag)} & \multicolumn{1}{c}{(mag)} \\ 
\hline
 1 & b & 3 & 15.25 & 13.20 & 12.04 &  1.84 & 14.90 \\
 4 & b & 4 & 16.94 & 13.51 & 11.70 &  2.88 & 14.90 \\
 5 & b & 7 &  \multicolumn{1}{c}{---}  & 15.94 & 13.93 &  3.23 & 15.17 \\
 7 & b & 29 & 13.99 & 12.09 & 11.01 &  1.69 & 14.97 \\
14 & b & 18 & 15.51 & 12.91 & 11.48 &  2.25 & 14.96 \\
16 & b & 19 &  \multicolumn{1}{c}{---}  & 14.62 & 12.30 &  3.70 & 14.59 \\
17 & b & 17 &  \multicolumn{1}{c}{---}  & 14.08 & 12.13 &  3.11 & 15.05 \\
18 & b & 21 & 17.54 & 14.27 & 12.25 &  3.23 & 14.94 \\
21 & b & 34 & 17.33 & 13.47 & 11.47 &  3.17 & 14.73 \\
23 & a & C1 & 19.13 & 14.76 & 12.71 &  3.27 & 15.27 \\
24 & a & C2 & 18.41 & 14.67 & 12.66 &  3.20 & 15.28 \\
\hline
\end{tabular}
\end{center}
\end{minipage}
\end{table}

\begin{figure}
\begin{minipage}{83mm}
\begin{center}
\includegraphics[clip,width=0.95\hsize]{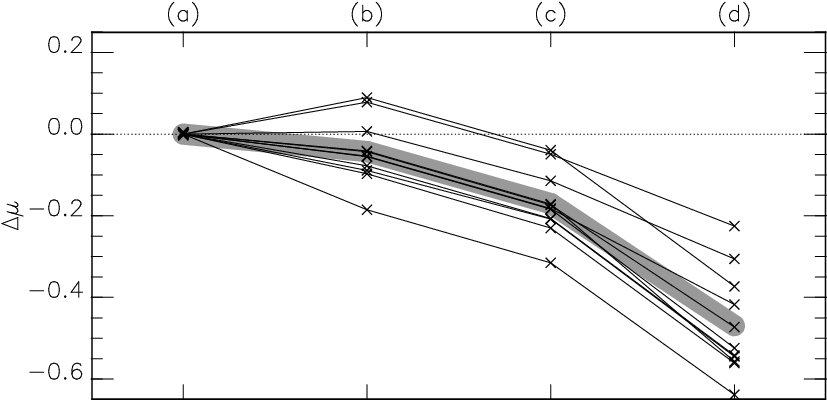}
\end{center}
\caption{
Distance moduli of 11 Cepheids, which were included in
both D\'{e}k\'{a}ny's work
and ours, are calculated with varying options of photometric data 
(either ours or those in \citealt{Dekany-2015a,Dekany-2015b}),
the PLR (\citealt{Matsunaga-2013} or \citealt{Monson-2012})
and extinction law (\citealt{Nishiyama-2006} or \citealt{Nishiyama-2009}),
and are compared with our values in Table~\ref{tab:catalogue}.
In case of (a), our photometric data were used with
our PLR (eqs.~\ref{eq:PHR} and \ref{eq:PKR}) and the extinction law
of \citet{Nishiyama-2006}, which is our default analysis,
and thus the offsets are zero.
D\'{e}k\'{a}ny's photometric data were used in case of (b).
In case of (c), furthermore, the PLR by \citealt{Monson-2012} were used.
The extinction law was further changed to the N09 for (d).
Grey thick line shows average offsets for the four cases.
\label{fig:DMdiff}}
\end{minipage}
\end{figure}

The differences in the distance moduli, $\Delta\mu$, can be attributed to
three reasons: measured magnitudes, the PLRs,
and the extinction law (Fig.~\ref{fig:DMdiff}).
First, magnitudes of classical Cepheids in Tables~\ref{tab:catalogue}
and \ref{tab:match} show significant scatter, $\sim$0.1~mag in $H$
and $\sim$0.05~mag in $\Ks$ when compared with each other.
This introduces a systematic offset of 0.04~mag
in distance moduli with a standard deviation of 0.08~mag,
which is illustrated by the (b) points in Fig.~\ref{fig:DMdiff}.
While the differences in the photometric results produce
the scatter of $\Delta\mu$, the systematic offset caused by this
is not large.

The second cause of $\Delta\mu$ is the difference in the PLRs adopted.
\citet{Dekany-2015a,Dekany-2015b} used PLRs by \citet{Monson-2012}
after the conversion of that photometric system to VISTA
and the re-calibration of their zero points based on
the distance to the LMC, 18.493~mag in $\mu$,
obtained by \citet{Pietrzynski-2013}.
Their PLRs are slightly different from the ones we
adopted (eqs.~\ref{eq:PHR} and \ref{eq:PKR}).
This leads to different estimates of
distance modulus by 0.07--0.1~mag depending on $\log P$ of a Cepheid
but independent of reddening; the PLRs by \citet{Monson-2012}
give smaller estimates of distance.
As indicated by the (c) points in Fig.~\ref{fig:DMdiff},
this increases the systematic offset of $\Delta\mu$ to 0.17~mag.

Finally, the difference in the extinction coefficients adopted is
the largest cause of the systematic offset.
With $\EHK$ values of 1--2~mag given, the difference of 0.17 in $\AKEHK$
between \citet{Nishiyama-2006} and \citet{Nishiyama-2009} introduces offsets
of 0.2--0.4~mag. Combining this with other offsets mentioned above,
we can explain the large systematic offsets, $\Delta\mu \sim 0.47$~mag.
This corresponds to $\sim$1.7~kpc at the distance of 8~kpc.

\begin{figure}
\begin{minipage}{83mm}
\begin{center}
\includegraphics[clip,width=0.95\hsize]{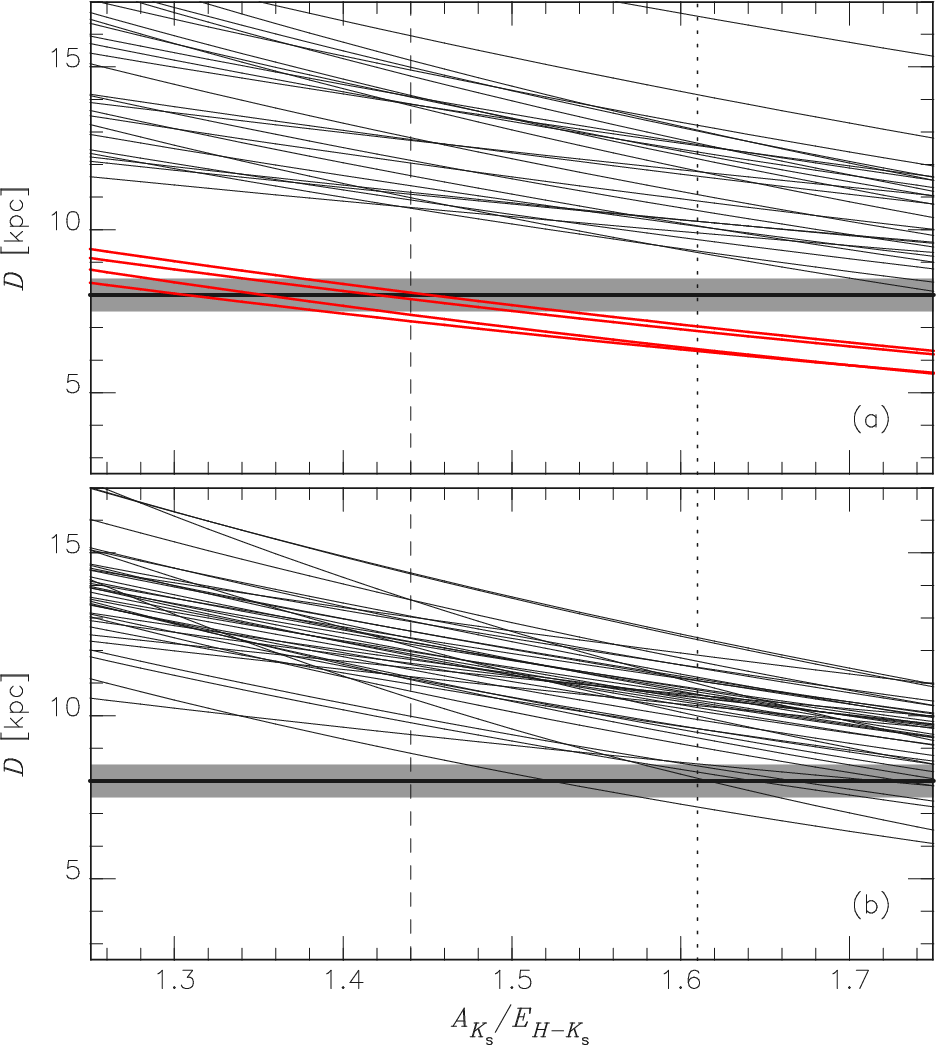}
\end{center}
\caption{Variation of estimated distances of Cepheids
plotted against the extinction coefficient $\AKEHK$.
The upper panel (a) is concerned with our Cepheids,
while the lower panel (b) with those in \citet{Dekany-2015a,Dekany-2015b}.
The vertical lines mark the coefficients from \citet{Nishiyama-2006,Nishiyama-2009}.
The horizontal line and grey area indicate the range of
recent estimates of the Galactic Centre distance, $8.0\pm 0.5$~kpc.
Four Cepheids which are considered to be within
the Nuclear Stellar Disk, having $D\sim 8$~kpc
with $\AKEHK \sim 1.44$, are indicated by red curves.
\label{fig:ED_HK}}
\end{minipage}
\end{figure}

Fig.~\ref{fig:ED_HK} illustrates the effect of $\AKEHK$ on
distances of Cepheids in this paper 
and those reported in \citet{Dekany-2015a,Dekany-2015b}.
The red curves in the upper panel corresponds to the results
of four Cepheids (Nos.~10--13 in Table~\ref{tab:catalogue}) which we consider
belong to the NSD.
The horizontal line and shaded region indicate
the recent estimate, $8.0 \pm 0.5$~kpc, of
the Galactic Centre distance \citep{Gillessen-2013}.
The comparison between the distances of the four Cepheids and
that of the Galactic Centre supports the coefficient of $\AKEHK = 1.44$
or even smaller. The larger coefficient $\AKEHK = 1.61$ would lead to
significantly smaller distances of the four Cepheids, $< 7$~kpc, which is
rather unexpected as the distance to the NSD
that is rotating around the Galactic Centre.
The membership of these four Cepheids in the NSD seems secure. This is based
on: (1) their positions, (2) their closely similar distances (independent
of the adopted reddening law), (3) their closely similar periods
(indicating an unusual period distribution and a significant increase
in star formation $\sim$25~Myr ago), and (4) their radial velocities
which are consistent with orbits expected in the NSD. Full details
of these aspects are given in \citet{Matsunaga-2011,Matsunaga-2015}.
Therefore, we should use the smaller $\AKEHK \sim 1.44$
at least for the four Cepheids towards the inner part of the bulge.
In addition, \citet{Schodel-2010} and \citet{Fritz-2011} estimated
$\AKEHK$ towards a small region around the Galactic Centre
to be 1.30--1.35, even smaller than our value,
which could put the average distance of the four NSD Cepheids
closer to 8.0~kpc.

How about other Cepheids distributed between $\pm 10$~degrees
in Galactic longitude?
The extinction law may be spatially variant.
\citet{Nishiyama-2006}, for example, suggested a variation
of $\sim$0.1 in $\AKEHK$ towards different corners of the inner bulge,
within 2.5~degrees around the Galactic Centre
(also see Gosling, Bandyopadhyay \& Blundell \citeyear{Gosling-2009};
\citealt{Chen-2013}; \citealt{Nataf-2013},
concerning the spatial variation of the near-IR extinction law).
If there were such variation,
this would be the largest source of uncertainty in distances of
the Cepheids discussed here. Nonetheless, if other Cepheids had
similar distances to the four Cepheids at $\sim$8~kpc, 
all of them should have large $\AKEHK$ values, 1.7 or more,
which would be distinctly large compared to the value for the NSD objects.
This is unlikely for the following reason.
Several previous papers traced the shape of the Galactic bar
using red clump distributed across a similar range of Galactic longitude,
$|l| < 10^\circ$ \citep{Nishiyama-2005,Rattenbury-2007,Wegg-2013}.
Using a given extinction coefficient for the entire range of each study,
the inclination of the bar appears as a smooth change of
the peak distance of the red clump.
If the $\AKEHK$ towards the Central region were significantly
different from other lines-of-sight, the bar structure
in the above studies should have presented the effect of
the systematic change instead of the smooth inclination.  
Therefore, we conclude that all the Cepheids in
this study and \citet{Dekany-2015a,Dekany-2015b} are 
further than the four Cepheids in the NSD.

\subsection{A lack of Cepheids in the inner part of the disk\label{sec:lackofCeps}}

\begin{figure}
\begin{minipage}{83mm}
\begin{center}
\includegraphics[clip,width=0.95\hsize]{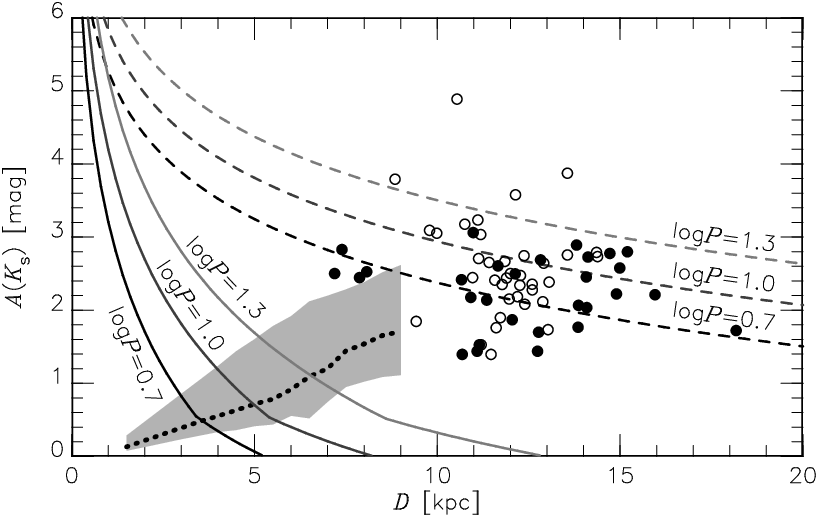}
\end{center}
\caption{
$D$ and $\AK$ of detected Cepheids are plotted---filled circles for
our objects and open circles for those in \citet{Dekany-2015a,Dekany-2015b}.
Note that the reddening law by \citet{Nishiyama-2006} was used for
both sample in this plot.
Also presented is the range of the $(D, \AK)$ space in which
classical Cepheids could be detected in both $H$- and $\Ks$-band
images in our survey.
Solid and dashed curves indicate saturation and fainter limits,
for each of which, curves corresponding to three different periods of Cepheids,
i.e.~different absolute magnitudes, are drawn.
Three-dimensional extinction map by \citet{Schultheis-2014},
$\AK$ increasing as a function of distance, is illustrated
by the thick dotted curve
and the shaded area; the former indicates the median curve of 
$\AK$ values along the lines-of-sight towards 142 IRSF/SIRIUS fields-of-view
in our survey
and the latter shows the range of 90 percentile at each distance.
\label{fig:DAKrangeCC}}
\end{minipage}
\end{figure}

In the following discussion, we use
$\AKEHK = 1.44$ for all of our Cepheids.
In addition, we re-calculated the distances of the Cepheids
reported by \citet{Dekany-2015a,Dekany-2015b} in the same way
as for our Cepheids using the PLRs of eqs.~(\ref{eq:PHR}) and (\ref{eq:PKR})
and the extinction law by \citet{Nishiyama-2006}.
The distances and extinctions obtained are plotted in
Fig.~\ref{fig:DAKrangeCC}.
The four Cepheids at around 8~kpc show
a concentration to the value expected for objects in the Galactic Centre,
while other objects show a wide range of $(D, \AK)$ values at $D>8$~kpc.
The Cepheids from \citet{Dekany-2015a,Dekany-2015b} are now largely
located behind the bulge. The revision of their distances does not
change their distances from the Galactic plane so much, and
they are within 50~pc of the plane except
one, Object~5 in \citet{Dekany-2015b}, being at $\sim$100~pc above the plane.
Their locations are thus consistent with the thin disk behind the bulge.

From the magnitude range of our survey (section \ref{sec:Observation})
and the expected absolute magnitude,
we can predict the range of $(D, \AK)$ in which we can detect Cepheids.
We here consider detections in both $H$ and $\Ks$, which are the
most important bands in this study.
Drawn in Fig.~\ref{fig:DAKrangeCC} are boundaries which correspond to
the detection limits (dashed curves) and the saturation limits (solid curves)
in the $(D, \AK)$ plane.
For each of the detection limits and the saturation limits,
three curves are drawn for Cepheids with different periods, i.e.~different
absolute magnitudes.
A Cepheid should have $\mu_0$ and $\AK$ which fall in between
the corresponding dashed and solid curves in Fig.~\ref{fig:DAKrangeCC}
in order to be detected in our survey.
With $\AK=2.5$, for example, Cepheids with $\log P=0.7$
are around the detection limit at the distance of $\sim$8~kpc,
while those with $\log P > 1.0$ can be detected
at a significantly larger distance.
It is therefore possible to detect longer-period Cepheids
within several kpc, both on the nearer and on the further side,
around the Galactic Centre unless the extinction gets
significantly larger.
The lack of our detection of such Cepheids indicates the absence
of Cepheids in the corresponding region.

We found no classical Cepheids significantly closer than
the four in the Galactic Centre although there seems to be room in
the parameter space in which Cepheids could have been detected in our survey.
The three-dimensional extinction map by \citet{Schultheis-2014} suggests
that the extinction remains smaller than 0.5~mag in $\AK$ up to
$\sim$3.5~kpc (Fig.~\ref{fig:DAKrangeCC}).
With such an extinction at the smaller distance, even short-period Cepheids
must have been saturated. No such Cepheids, if any, in our survey field
have been reported in previous variability catalogues
(DDO Database of Galactic Classical Cepheids, \citealt{Fernie-1995}; General Catalogue of Variable Stars, \citealt{Samus-2015}; AAVSO International Variable Star Index, Watson, Henden \& Price \citeyear{Watson-2015}).

In the range of 3.5--6~kpc in distance, some Cepheids could have been
detected in our survey depending on the period
and the foreground extinction. This range corresponds to the Galactocentric
distance of 2--4.5~kpc with the Galactic Centre distance assumed to be 8~kpc,
thus in the vicinity of the 3~kpc arm and closer to the Centre.
Our survey field, $\sim$2.3~deg$^2$,
covers 0.04~kpc$^{3}$ in this distance range along the Galactic mid-plane. 
The DDO Database \citep{Fernie-1995} lists 71 Cepheids within
2~kpc around the Sun and within 50~pc of the Galactic plane,
which gives a volume density of 110~kpc$^{-3}$.
Assuming the density gets higher towards the inner region
following an exponential law with a scale length of 3.5~kpc
(see Windmark, Lindegren \& Hobbs \citeyear{Windmark-2011}),
the density in the above volume
would be higher by a factor of $\sim$3, thus leading to
$\sim$350~kpc$^{-3}$. This suggests that about 14 Cepheids could be
included within our survey field at 3.5--6~kpc, although we found none.

In order to take the completeness of our survey into account, 
we made a Monte-Carlo simulation as follows.
First, we generate 5000 Cepheids uniformly distributed within
our survey volume between 3.5--6~kpc.
Then, we assign their periods randomly but following the period distribution
of the DDO Database Cepheids in
the solar neighbourhood (see Fig.~2 of \citealt{Matsunaga-2011}).
With the PLRs (eqs.~\ref{eq:PHR} and \ref{eq:PKR})
and the extinction from the map by \citet{Schultheis-2014}
for each simulated Cepheid,
we can calculate its expected magnitudes
in $H$ and $\Ks$ and determine whether it would be detected
in our survey or not. The simulation suggests that $\sim$70 percents of
Cepheids located in this volume should have been detected.
Considering the space density mentioned above, we could have detected about 10
Cepheids in our survey with the assumed space distribution.
The absence of such Cepheids indicates that the density of Cepheids
in the inner part of the disk, except the NSD, is
lower than expected by the simple exponential law.
The volume density in the inner part may be even smaller than
that in the solar neighbourhood with which $\sim$3 Cepheids are expected to
be found in our survey.

\subsection{Distribution of Cepheids across the Galactic plane\label{sec:Distribution}}

\begin{figure*}
\begin{minipage}{150mm}
\begin{center}
\includegraphics[clip,width=0.90\hsize]{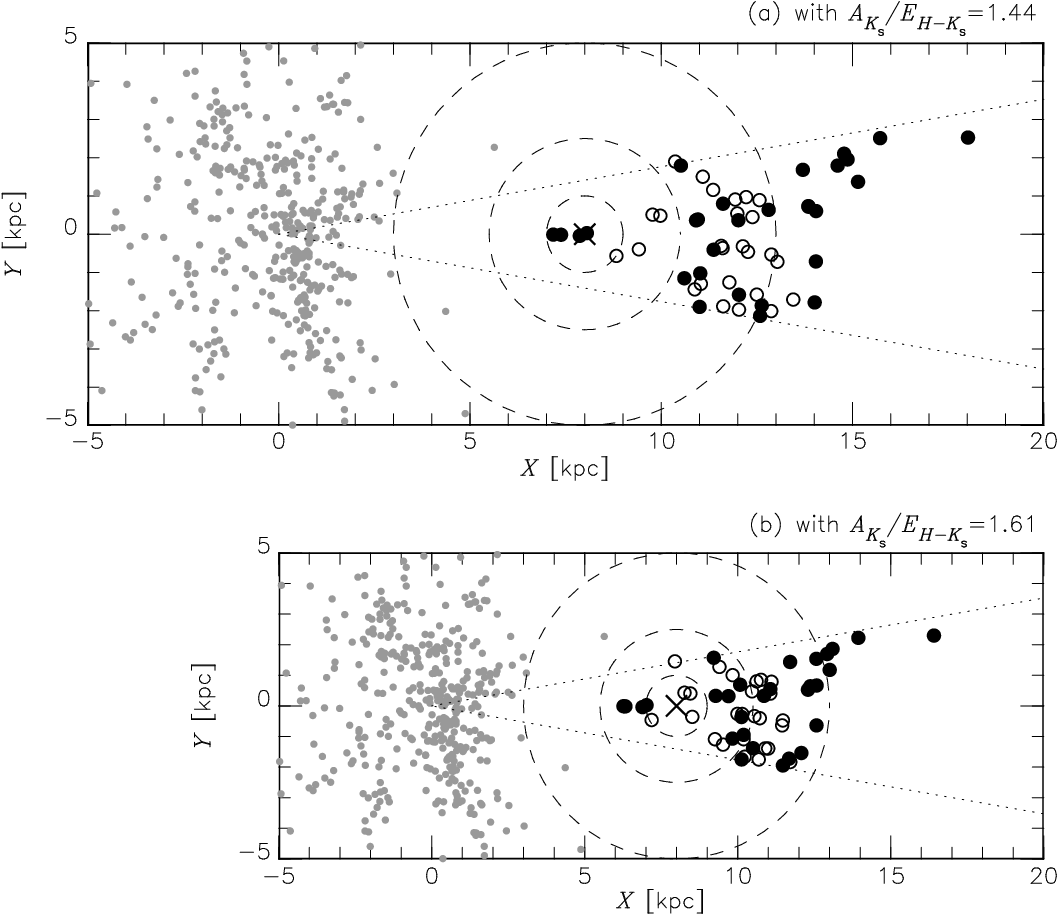}
\caption{
Distribution of Cepheids, ours and those in \citet{Dekany-2015a,Dekany-2015b}
indicated by filled and open circles, respectively, on the face-on view of
the Galactic disk. 
The reddening law by \citet{Nishiyama-2006} was used
for both sample in the panel (a), while that by \citet{Nishiyama-2009}
was used in the panel (b).
Note the displacement of the four NSD Cepheids from the Centre in (b).
If a Cepheid is found in both surveys,
a distance obtained with our photometry is used.
Grey points show the distribution of previous Cepheids
in \citet{Fernie-1995}. The Sun is located at the origin, and
the Galactic Centre
is indicated by the cross at an assumed distance of 8~kpc.
Dashed circles at 1, 2.5 and 5~kpc
from the Galactic Centre are drawn for readers' convenience.
The longitude range of our survey, $|l|<10^\circ$, is illustrated by
dotted lines.
\label{fig:XYsCC}}
\end{center}
\end{minipage}
\end{figure*}

Fig.~\ref{fig:XYsCC} plots the distribution of
our Cepheids and D\'{e}k\'{a}ny's sample by filled and open circles.
Locations of the DDO Database Cepheids are indicated by grey points
(e.g.\  see Majaess, Turner \& Lane, \citeyear{Majaess-2009},
for discussions on their distribution).
Here, again, the distances of the D\'{e}k\'{a}ny's objects are
obtained based on our adopted PLR and the extinction law of \citet{Nishiyama-2006}.
Except the four Cepheids belonging to the NSD,
none of our Cepheids, indicated by filled circles,
are within 2.5~kpc around the Galactic Centre
indicated by the cross. 
Therefore, we found no evidence of the disk-like distribution of
young stars in the inner part of the Galaxy
suggested by \citet{Dekany-2015b}, except the NSD.

Towards the bulge but in regions of higher Galactic latitude, $|b|>2^\circ$,
\citet{Soszynski-2011} reported 32 classical Cepheids
found in the Optical Gravitational Lensing Experiment (OGLE).
The apparent magnitudes of these Cepheids are similar to 
those of type II Cepheids in the bulge (fig.~7 in \citealt{Soszynski-2011}),
and thus, if these are indeed classical Cepheids,
they are expected to be significantly further than the bulge.
Five of them are confirmed to be classical Cepheids located in
the flared part of the disk by \citet{Feast-2014},
while accurate distances and natures of the others are
not well determined.
The five Cepheids in \citet{Feast-2014} are located outside the range
of Figure~\ref{fig:XYsCC}, at heliocentric distances of 22~kpc or larger.
Their distances from the Galactic plane,
$> 1$~kpc, are also distinctly larger than the Cepheids plotted
in Figure~\ref{fig:XYsCC}.

Among the Cepheids reported by D\'{e}k\'{a}ny's {et~al.},
four stars (their 
11, 13, 20, and 22) are relatively close to the Galactic Centre,
1--2~kpc. Unfortunately, none of them are within our survey field. 
Their exact locations and kinematics are of great interest
to study stellar populations of the inner part of the Galaxy.

On the other hand, as we discussed in Section~\ref{sec:lackofCeps},
we found no Cepheids closer than the four immediately
around the Galactic Centre. In addition, there seems to be a 
space, at the Galactic distance of 2.5--5~kpc, without any Cepheids
on the nearer side of the Galactic Centre in contrast to
the further side where dozens Cepheids were found
(\citealt{Dekany-2015a,Dekany-2015b} and this work).
Our survey only partly covers
this relatively close range (3--5.5~kpc from the Sun)
due to the saturation limit (Fig.~\ref{fig:DAKrangeCC}).
Previous surveys may also have been incomplete in this region
since our survey and that of \citet{Dekany-2015a,Dekany-2015b}
have found Cepheids at the corresponding Galactocentric distance
on the opposite side of the Centre.
Deeper and more comprehensive surveys in 
the optical, such as {\it Gaia}, or shallower surveys in the infrared
will have the capability to provide
a more complete mapping of variable star populations.

\section{Concluding remarks\label{Summary}}

The presence of an inner thin disk,
represented by Cepheids, which was
suggested by \citet{Dekany-2015b} has not been confirmed.
The lack of Cepheids in the inner part of the Galaxy, 
a few kilo-parsecs both on the front- and far-side of the Galactic Centre,
suggests that the young stellar populations do not follow
the exponential disk distribution towards the Centre.
A similar lack in the inner disk has been also found
in the distribution of H$_{\rm II}$ regions.
Our study demonstrated that 
the extinction law has a strong impact on investigations
of the distribution of obscured stars in the inner Galaxy.
It is nonetheless clear that the recent infrared surveys are
revealing populations of Cepheids in a large area
of the Galaxy, including the opposite side of the disk
beyond the bulge. Detailed observations of these objects,
such as high-resolution spectroscopy for radial velocities
and metallicities, would provide a new path to
a global picture of Galactic structure and evolution.

\section*{Acknowledgments}

We are grateful to Naoko Asami, Hirofumi Hatano, Nobuyuki Ienaka,
Nobuhiko Kusakabe, Nagisa Oi and Ihab Riad
who contributed to our monitoring observations at the IRSF.
This work has been supported by Grants-in-Aid 
(Nos.~07J05097, 19740111, 23684005, 26287028) from the
Japan Society for the Promotion of Science (JSPS).
MWF acknowledges support from the National Research Foundation (NRF)
of South Africa.
GB acknowledges the JSPS invitation fellowship which enabled him
to stay in Japan and to have active collaborations with the host researcher,
NM, and other Japanese colleagues.
LI appreciates funding support from the Sonderforschungsbereich SFB 881
``The Milky Way System'' (subproject C9) of
the German Research Foundation (DFG).

\bibliographystyle{mn2e}

\begin{thebibliography}{99}
\bibitem[\protect\citeauthoryear{Anderson {et~al.}}{2011}]{Anderson-2011}
Anderson L.~D., Bania T.~M., Balser D.~S., Rood R.~T., 2011, ApJS, 194, 32
\bibitem[\protect\citeauthoryear{Anderson {et~al.}}{2012}]{Anderson-2012}
Anderson L.~D., Bania T.~M., Balser D.~S., Rood R.~T., 2012, ApJ, 754, 62
\bibitem[\protect\citeauthoryear{Benedict {et~al.}}{2007}]{Benedict-2007}
Benedict G.~F. et al., 2007, AJ, 133, 1810
\bibitem[\protect\citeauthoryear{Bono {et~al.}}{2005}]{Bono-2005}
Bono G., Marconi M., Cassisi S., Caputo F., Gieren W., Pietrzynski G., 2005, ApJ, 621, 966
\bibitem[\protect\citeauthoryear{Chen {et~al.}}{2013}]{Chen-2013}
Chen B.~Q., Schultheis M., Jiang B.~W., Gonzalez O.~A., Robin A.~C., Rejkuba M., Minniti D., 2013, A\&A, 550, A42
\bibitem[\protect\citeauthoryear{Dame \& Thaddeus}{2008}]{Dame-2008}
Dame T.~M. \& Thaddeus P., 2008, ApJ, 683, L143
\bibitem[\protect\citeauthoryear{D\'{e}k\'{a}ny {et~al.}}{2015a}]{Dekany-2015a}
D\'{e}k\'{a}ny I. et~al., 2015a, ApJ, 799, L11
\bibitem[\protect\citeauthoryear{D\'{e}k\'{a}ny {et~al.}}{2015b}]{Dekany-2015b}
D\'{e}k\'{a}ny I. et~al., 2015b, ApJ, 812, L29
\bibitem[\protect\citeauthoryear{Feast {et~al.}}{2014}]{Feast-2014}
Feast M.~W., Menzies J.~W, Matsunaga N., Whitelock P.~A., 2014, Nat, 509, 342
\bibitem[\protect\citeauthoryear{Fernie {et~al.}}{1995}]{Fernie-1995}
Fernie J.~D., Evans N.~R., Beattie B., Seager S., 1995, IBVS, 4148, 1
\bibitem[\protect\citeauthoryear{Figer {et~al.}}{1999}]{Figer-1999}
Figer D.~F., McLean I.~S., Morris M., 1999, ApJ, 514, 202
\bibitem[\protect\citeauthoryear{Figer {et~al.}}{2002}]{Figer-2002}
Figer D.~F. et al., 2002, ApJ, 581, 258
\bibitem[\protect\citeauthoryear{Fritz {et~al.}}{2011}]{Fritz-2011}
Fritz T.~K. et al., 2011, ApJ, 737, 73
\bibitem[\protect\citeauthoryear{Gillessen {et~al.}}{2013}]{Gillessen-2013}
Gillessen S., Eisenhauer F., Fritz T.~K., Pfuhl O., Ott T., Genzel R., 2013, IAUS, 289, 29
\bibitem[\protect\citeauthoryear{Gosling {et~al.}}{2009}]{Gosling-2009}
Gosling A.~J., Bandyopadhyay R.~M., Blundell K.~M., 2009, MNRAS, 394, 2247
\bibitem[\protect\citeauthoryear{Jones {et~al.}}{2013}]{Jones-2013}
Jones C., Dickey J.~M., Dawson J.~R., McClure-Griffiths N.~M.,
Anderson L.~D., Bania T.~M., 2013, ApJ, 774, 117
\bibitem[\protect\citeauthoryear{Launhardt {et~al.}}{2002}]{Launhardt-2002}
Launhardt R., Zylka R., Mezger P.~G., 2002, A\&A, 384, 112
\bibitem[\protect\citeauthoryear{Majaess {et~al.}}{2009}]{Majaess-2009}
Majaess D.~J., Turner D.~G., Lane D.~J., 2009, MNRAS, 398, 263
\bibitem[\protect\citeauthoryear{Matsunaga {et~al.}}{2009}]{Matsunaga-2009}
Matsunaga N., Kawadu T., Nishiyama S., Nagayama T., Hatano H.,
Tamura M., Glass I.~S., Nagata T., 2009, MNRAS, 399, 1709
\bibitem[\protect\citeauthoryear{Matsunaga {et~al.}}{2011}]{Matsunaga-2011}
Matsunaga N. et al., 2011, Nat, 477, 188
\bibitem[\protect\citeauthoryear{Matsunaga {et~al.}}{2013}]{Matsunaga-2013}
Matsunaga N. et al., 2013, MNRAS, 429, 385
\bibitem[\protect\citeauthoryear{Matsunaga}{2014}]{Matsunaga-2014}
Matsunaga N., 2014, EAS Pub.~Ser., 67, 279
\bibitem[\protect\citeauthoryear{Matsunaga {et~al.}}{2015}]{Matsunaga-2015}
Matsunaga N. et al., 2015, ApJ, 799, 46
\bibitem[\protect\citeauthoryear{Mauerhan {et~al.}}{2010}]{Mauerhan-2010}
Mauerhan J.~C., Cotera A., Dong H., Morris M.~R., Wang Q.~D., Stolovy S.~R., Lang C., 2010, ApJ, 725, 188
\bibitem[\protect\citeauthoryear{Minniti {et~al.}}{2010}]{Minniti-2010}
Minniti D. et al., 2010, New Astron., 15, 433
\bibitem[\protect\citeauthoryear{Monson {et~al.}}{2012}]{Monson-2012}
Monson A.~J., Fredman W.~L., Madore B.~F., Persson S.~E., Scowcroft V., Seibert M., Rigby J.~R., 2012, ApJ, 759, 146
\bibitem[\protect\citeauthoryear{Nagashima {et~al.}}{1999}]{Nagashima-1999}
Nagashima C. et al., 1999, in Nakamoto T., ed, Proc. Star Formation 1999. Nobeyama Radio Observatory, Nagano, p.~397
\bibitem[\protect\citeauthoryear{Nagayama {et~al.}}{2003}]{Nagayama-2003}
Nagayama T. et al., 2003, in Iye M., Moorwood A.~F.~M., eds, SPIE Vol.~4841,
\bibitem[\protect\citeauthoryear{Nataf {et~al.}}{2013}]{Nataf-2013}
Nataf D.~M. et al., 2013, ApJ, 769, 88
\bibitem[\protect\citeauthoryear{Nishiyama {et~al.}}{2005}]{Nishiyama-2005}
Nishiyama S. et al., 2005, ApJ, 621, L105
\bibitem[\protect\citeauthoryear{Nishiyama {et~al.}}{2006}]{Nishiyama-2006}
Nishiyama S. et al., 2006, ApJ, 638, 839
\bibitem[\protect\citeauthoryear{Nishiyama {et~al.}}{2009}]{Nishiyama-2009}
Nishiyama S., Tamura M., Hatano H., Kato D., Tanab\'{e} T., Sugitani K., Nagata T., 2009, ApJ, 696, 1407
\bibitem[\protect\citeauthoryear{Pietrzy\'{n}ski {et~al.}}{2013}]{Pietrzynski-2013}
Pietrzy\'{n}ski G. et al., 2013, Nat, 495, 76
\bibitem[\protect\citeauthoryear{Rattenbury {et~al.}}{2007}]{Rattenbury-2007}
Rattenbury N.~J., Mao S., Sumi T., Smith M.~C., 2007, MNRAS, 378, 1064
\bibitem[\protect\citeauthoryear{Saito {et~al.}}{2011}]{Saito-2011}
Saito R.~K., Zoccali M., McWilliam A., Minniti D., Gonzalez O.~A., Hill V., 2011, AJ, 142, 76
\bibitem[\protect\citeauthoryear{Samus {et~al.}}{2015}]{Samus-2015}
Samus N.~N., Durlevich O.~V., Goranskij V.~P., Kazarovets E.~V., Kireeva N.~N., Pastukhova E.~N., Zharova A.~V., 2015, General Catalogue of Variable Stars
\bibitem[\protect\citeauthoryear{Sanna {et~al.}}{2014}]{Sanna-2014}
Sanna A. {et~al.}, 2014, ApJ, 781, 108
\bibitem[\protect\citeauthoryear{Schultheis {et~al.}}{2014}]{Schultheis-2014}
Schultheis M. {et~al.}, 2014, A\&A, 566, A120
\bibitem[\protect\citeauthoryear{Sch\"{o}del {et~al.}}{2010}]{Schodel-2010}
Sch\"{o}del R., Najarro F., Muzic K., Eckart A., 2010, A\&A, 511, A18
\bibitem[\protect\citeauthoryear{Soszy\'{n}ski {et~al.}}{2011}]{Soszynski-2011}
Soszy\'{n}ski I. {et~al.}, 2011, Acta Astron., 61, 285
\bibitem[\protect\citeauthoryear{van~Leeuwen {et~al.}}{2007}]{vanLeeuwen-2007}
van~Leeuwen F., Feast M.~W., Whitelock P.~A., Laney C.~D., 2007, MNRAS, 379, 723
\bibitem[\protect\citeauthoryear{Watson {et~al.}}{2015}]{Watson-2015}
Watson C., Henden A.~A., Price A., 2015, AAVSO International Variable Star Index
\bibitem[\protect\citeauthoryear{Wegg \& Gerhard}{2013}]{Wegg-2013}
Wegg C., Gerhard O., 2013, MNRAS, 435, 1874
\bibitem[\protect\citeauthoryear{Windmark {et~al.}}{2011}]{Windmark-2011}
Windmark F., Lindegren L., Hobbs D., 2011, A\&A, 530, A76
\bibitem[\protect\citeauthoryear{Yusef-Zadeh {et~al.}}{2009}]{YusefZadeh-2009}
Yusef-Zadeh F. {et~al.}, 2009, ApJ, 702, 178
\end{thebibliography}

\label{lastpage}

\end{document}